\documentclass[12pt,epsf,epsfig,psfig]{article}
\usepackage{graphicx}
\usepackage{epstopdf}
\usepackage{epsfig}
\usepackage{cite}
\def\e{\epsilon}

\def\be{\begin{equation}}
\def\ee{\end{equation}}

\def\lsim{\raise0.3ex\hbox{$<$\kern-0.75em\raise-1.1ex\hbox{$\sim$}}}
\def\gsim{\raise0.3ex\hbox{$>$\kern-0.75em\raise-1.1ex\hbox{$\sim$}}}

\def\NP{{ Nucl.\ Phys.\ }}
\def\PL{{ Phys.\ Lett.\ }}
\def\PR{{ Phys.\ Rev.\ }}

\def\PRL{{ Phys.\ Rev.\ Lett.\ }}

\def\ZP{{ Z.\ Phys.\ }}
\def\EP{{ Europ.\ Phys.\ J.\ C}}

\begin{document}


~ \hfill BI-TP 2016/02(rev)

~~\vskip0.7cm

\centerline{\Large \bf Universal Strangeness Production}

\vskip0.3cm

\centerline{\Large \bf in Hadronic and Nuclear Collisions}

\vskip0.6cm 

\centerline{\bf P.\ Castorina$^{\rm a,b}$, S.\ Plumari$^{\rm a,c}$
and H.\ Satz$^{\rm d}$} 

\bigskip

\centerline{a: Dipartimento di Fisica ed Astronomia, 
Universit\'a di Catania, Italy}

\centerline{b: INFN, Sezione di Catania, Catania, Italy}

\centerline{c: INFN-LNS, Catania, Italy} 

\centerline{d: Fakult\"at f\"ur Physik, Universit\"at Bielefeld, Germany}



\vskip1cm

\centerline{\large \bf Abstract}

\bigskip

We show that strangeness suppression in hadronic and nuclear collisions is 
fully determined by the initial energy density of the collision. The
suppression factor $\gamma_s(s)$, with $\sqrt s$ denoting the collision 
energy, can be expressed as a universal function of the initial energy density
$\e_0(s)$, and the resulting pattern is in excellent agreement with
data from $p-p,~p-Pb,~Cu-Cu,~Au-Au$ and $Pb-Pb$ data over a wide range of 
energies and for different centralities.

\vskip1cm

\begin{figure}[h]
\centerline{\psfig{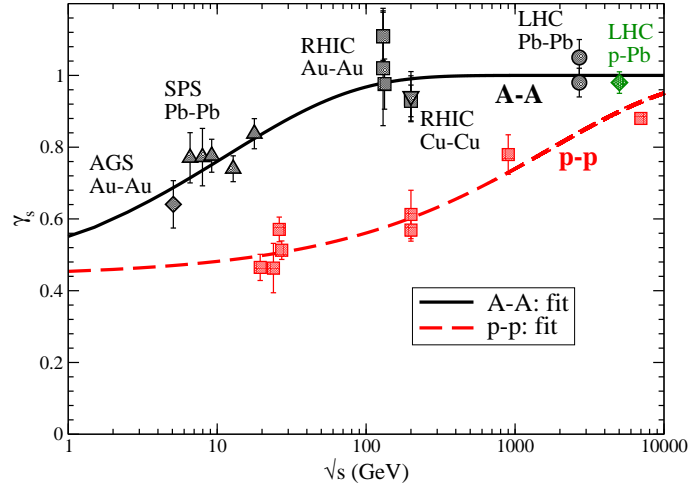}}
\caption{The strangeness suppression factor $\gamma_s$ as function of the
collision energy $\sqrt s$ for $pp$ (red symbols), $Pb-Pb$, $Au-Au$, 
$Cu-Cu$ (black symbols) and $p-Pb$ (green symbol) collisions \cite{data1}.
}
\label{gamma}
\end{figure}

The relative hadron production rates in high energy strong interactions are 
in general well accounted for in terms of an ideal resonance gas at
temperature $T$ and baryochemical potential $\mu$ , with one
significant caveat. Over a wide range of collision energies, both in $pp$ 
and in $AA$ collisions, the rates for the production of strange hadrons 
are found fall below the values predicted by an ideal resonance gas in 
chemical equilibrium; see e.g. \cite{Beca-Passa,PBM-R-St,Beca-LB}.  
The resonance gas scenario can be maintained, however, 
by the rather {\sl ad hoc} introduction of a strangeness suppression factor 
$\gamma_s(s) < 1$, with $\gamma_s^n$ reducing the production rate of
hadrons containing $n$ strange quarks or antiquarks \cite{Rafelski}.
With increasing
collision energies $\sqrt s$, the strangeness suppression decreases, i.e., 
$\gamma_s(s)$ is found to approach unity. In Fig.\ \ref{gamma}, we show 
the behavior of $\gamma_s(s)$ as function of the collision energy $\sqrt s$ 
in $pp$ and heavy ion ($Pb-Pb$, $Au-Au$, $Cu-Cu$) collisions \cite{data1}; 
we return to the
details of the mentioned fits a little later on. With such a suppression
factor, one finds excellent agreement for the relative hadron abundances
in hadronic and nuclear collisions as well as in $e^+e^-$ annihilation. 

\medskip

Nevertheless, the origin and functional behavior of $\gamma_s$ has for a
long time remained enigmatic, as has the difference in behavior between
elementary and nuclear collisions. A first step towards a solution was 
obtained by noting that in collisions producing only a small number of
strange particles, strangeness conservation should be enforced not only 
exactly (canonical instead of grand canonical) \cite{HR}, but moreover
on a local
level, within a strangeness correlation volume $V_c < V$: the production of 
a single strange particle would require that of an antiparticle nearby, not 
somewhere in some large equivalent global volume $V$ \cite{HRT}.
Such a requirement is effectively a deviation from global ideal gas behavior,
and it seems necessary only if there are just a few strange hadrons. Once
their numbers become large enough, local compensation is automatically
given, so that there is no longer any need for a specific strangeness 
correlation volume. And indeed one finds that for $V_c/V \to 1$, the
corresponding resonance gas 
predictions converge to those of an equilibrium grand canonical formulation.

\medskip

The theoretical basis for such strangeness correlation volumes was recently
provided through causality considerations for the space-time evolution of
high energy interactions \cite{CS1,CS2}.
In a boost-invariant production scenario \cite{Bjorken}, one has after a
brief thermalization stage an intermediate thermal medium in strong 
interaction; this then freezes out into free hadrons. 
At the thermalization time, all information about the initial state is
lost, and beyond the freeze-out time, there is no further interaction
between the individual hadrons. In \cite{CS1,CS2} it was shown that 
between these two times, the strongly interacting thermal medium, whether 
deconfined (QGP) or confined (interacting hadrons), is partioned into 
causally disconnected space-time regions, similar to the horizon problem
in cosmology, with no communication possible between different regions.
Hadrons produced at large rapidity come from a thermal fireball
which is causally disjoint from a fireball leading to low 
rapidity hadrons, and so one cannot expect strangeness conservation 
to occur through interaction between the relevant bubbles. The concept of 
a global equivalent cluster \cite{Beca-Passa}
thus cannot be applied here: exactly conserved quantum 
numbers have to be conserved within (smaller) causally connected volumes.
This implies that any dynamical correlations
among regions separated by a large rapidity gap must originate before
the equilibration time.

\medskip

\begin{figure}[h]
\centerline{\psfig{file=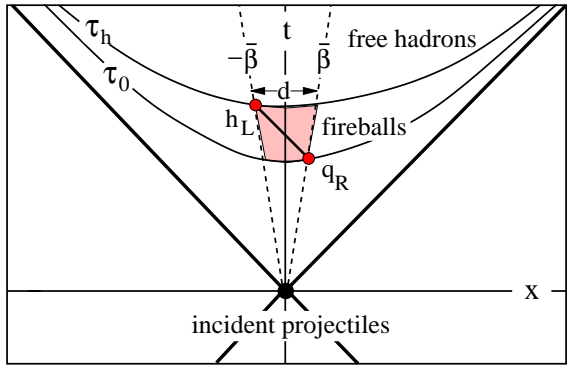,width=7.5cm}}
\caption{The formation and evolution of a fireball at rest in the
center of mass; the fireball is indicated in pink.}
\label{evo}
\end{figure}

The aim of the present paper is to show that this scenario leads to a
universal description of strangeness production in high energy collisions,
providing a common formulation for $pp$ and $AA$ collisions at different
collision energies, different centralities, and different $A$. In Fig.\
\ref{evo}, we illustrate the definition of a fireball in terms of the
collision evolution: we require a causal connection between the most
separate points ($q_R$ and $h_L$) of the bubble. 

\medskip

In such a scheme, the size $d$ of the causal correlation region is
determined by the values of the thermalization time $\tau_0$ (the
``equilibration'' time) and the hadronisation time $\tau_h$, specifying
the emission of free hadrons. Geometric considerations \cite{CS1}
result in 
\be
{d \over \tau_0} = \sqrt{\frac{\tau_h}{\tau_0}}
\left({\tau_h \over \tau_0} - 1\right).
\label{1}
\ee
For boost-invariant hadronisation, the time evolution is governed by the
one-dimensional hydrodynamic expansion
\be
\frac{d\epsilon}{d\tau}= - {(\epsilon+p) \over \tau},
\label{2}
\ee
where $\e$ denotes the energy density and $p$ the pressure.
Given the equation of state of the plasma, we can thus express $\tau$ as
function of the energy density $\e$. To illustrate, we consider an ideal
quark-gluon plasma, with $p=\e / 3$, to obtain
\be
\frac{\tau_h}{\tau_0}=\left(\frac{\epsilon_0}{\epsilon_h}\right)^{3/4}.
\label{3}
\ee
As other extreme, we set $p=0$, to find
\be
\frac{\tau_h}{\tau_0}=\left(\frac{\epsilon_0}{\epsilon_h}\right).
\label{4}
\ee
For a realistic description, we can take the equation of state as determined
in finite temperature lattice QCD studies 
\cite{lattice-eos1,lattice-eos2}, which with 
$p=a \e,~0<a<1/3$, gives us
\be
\frac{\tau_h}{\tau_0}=\left(\frac{\epsilon_0}{\epsilon_h}\right)^{1/(1+a)}.
\label{5}
\ee
a value lying somewhere between the two extremes just considered. For our
present considerations, however, the precise form is not important, as we
shall see.

\medskip

What is crucial is that the spatial size $d$ of the causally correlated 
region is determined by $\e_0/\e_h$. Since the universal hadronisation 
energy density $\e_h \simeq 0.4 - 0.6$ GeV/fm$^3$ is obtained in lattice 
QCD studies and $\tau_0$ is conventionally assumed to be about 1 fm,  
we find that the correlation volume is effectively specified by the initial 
energy density $\e_0(s)$. For a central collision in the boost-invariant 
scheme assumed here, this is given by the Bjorken expression \cite{Bj} 
\be 
\e_0 ~\!\tau_0 =
{1.5~\!A \over \pi R_A^2} \left({dE \over dy}
\right)_{y=0}^{AA},
\label{6}
\ee
with $R_A = 1.25 A^{1/3}$. Here $A(dE/dy)_{y=0}^{AA}$ is the average energy
deposited in the $y=0$ interval of a central $AA$ collision; hence
$(dE/dy)_{y=0}^{AA}$ denotes the average energy per one-half the 
number of participants in the nuclear interaction volume. 
A simplistic approximation gives
\be
\left({dE \over dy}\right)_{y=0}^{AA} \simeq m_T\left({dN \over dy}
\right)_{y=0}^{AA},
\label{6a}
\ee
where $m_T \simeq 0.5$ GeV is the average transverse energy per produced
hadron and $({dN/dy})_{y=0}^{AA}$ the average multiplicity
per one-half the number of participants deposited in the nuclear
interaction volume. The result is the often-used form
\be 
\e_0 ~\!\tau_0 =
{1.5~\! m_T A \over \pi R_A^2} \left({dN \over dy}
\right)_{y=0}^{AA},
\label{6b}
\ee
for the central energy density. It effectively ignores the energy density
component arising from hydrodynamic flow; nevertheless, we shall use it
to begin with and later on return to the modification arising when flow
is included.

\medskip

The average charged
multiplicity in $Pb-Pb$ and $Au-Au$ collisions at central rapidity and
per one-half the number of participants has been parametrized by \cite{mult} 
\be
\left({dN \over dy}\right)_{y=0}^{AA} = a(\sqrt s)^{0.3} + b,
\label{7}
\ee
with a=0.7613 and b= 0.0534 ( $A(dN/dy)^{AA}_{y=0}$ is the overall
charged multiplictiy in a central $AA$ collision).
The counterpart for $pp$ collisions is
\be 
\e_0^p~\!\tau_0 = {1.5~\! m_T \over \pi R_p^2} \left({dN \over dy}
\right)_{y=0}^{pp},
\label{8}
\ee
with
\be
\left({dN \over dy}\right)_{y=0}^{pp} = a(\sqrt s)^{0.22} + b,
\label{9}
\ee
where $R_p = 0.8$ fm, a=0.797 and b= 0.04123 \cite{mult}.

\medskip

These expressions specify the initial energy density $\e_0$ in terms
of the collision energy $\sqrt s$, and this in turn gives us the causal
correlation size $d$. It is thus the initial energy density which 
determines the degree of strangeness suppression. If this is correct,
we should be able to relate directly the strangeness suppression as
given by $\gamma_s(s)$ to the initial energy densities of $pp$ and
$AA$ collisions.

\medskip

\begin{figure}[h]
\centerline{\psfig{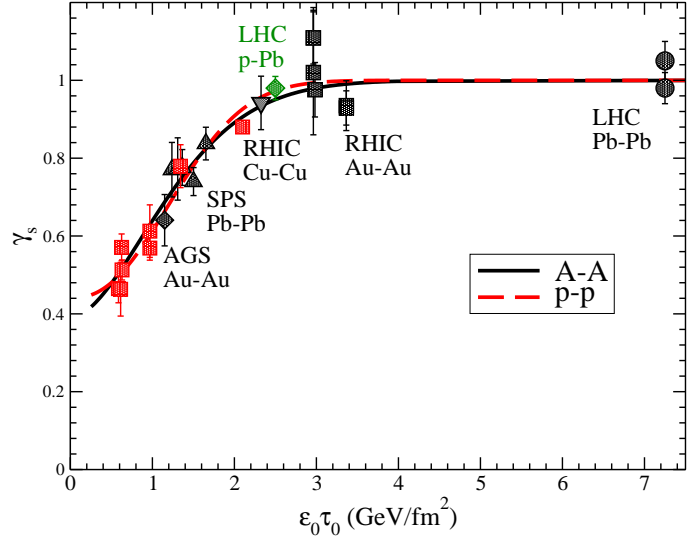}}
\caption{Strangeness suppression as function of the initial energy density
in $pp$, $pPb$ and $AA$ collisions, based on eq.(8); black symbols 
correspond to $AA$, green to $pPb$ and red to $pp$ data.}
\label{supp}
\end{figure}

To test this, we use the fits to $\gamma_s(s)$ given in \cite{beca,plu},
\be
\gamma_s^A(s) = 1 - a_A \exp{(-b_A \sqrt{A \sqrt s})}
\label{10}
\ee
and
\be
\gamma_s^p(s) = 1 - a_p \exp{(-b_p s^{1/4}}),
\label{11}
\ee
with $a_A=0.606,~a_p=0.5595,~b_A=0.0209,~b_p=0.0242$. In Fig.\ \ref{gamma}, 
these fits were shown together with high energy data. At a given collision
energy $s$, we thus have from equ'ns.\ (\ref{6}-\ref{9}) the average
energy density $\epsilon_0(s)$ and from equ'ns.\ (\ref{10} - \ref{11})
the corresponding value of $\gamma_s(s)$. As a result, we therefore obtain
$\gamma_s(s)$ as a function of $\epsilon_0(s)$. The result
is shown in Fig.\ \ref{supp} together with the available data from SPS
to LHC energies.

\medskip

In Fig.\ \ref{gamma}, we had also included the LHC point for $p-Pb$
collisions at $\sqrt s = 5.02$ TeV. To determine the corresponding 
energy density, we use in eq.\ (\ref{6}) 
\be
R_T = R_p (0.5 {\bar N}_{\rm part})^{1/3}
\label{pA}
\ee
for the transverse radius, with ${\bar N}_{\rm part}\simeq 8$, as given by
\cite{Abelev}, together with the average secondary multiplicity found
there. The resulting point is included in Fig.\ \ref{supp}.

\medskip

In Fig.\ \ref{supp},
we note first of all that the functional forms of $\gamma_s$ for $AA$ and 
$pp$ collisions in terms of $\e_0(s)\tau_0$ fully coincide; the difference
between elementary and nuclear collisions seen in Fig.\ \ref{gamma} is simply 
due to the fact that strangeness suppression is not determined by the 
overall collision energy. Instead, it is governed by the initial energy 
density. Next we note that in fact all data are in excellent agreement with
such a universal energy density scaling of strangeness suppression. 

\medskip

\begin{figure}[h]
\centerline{\psfig{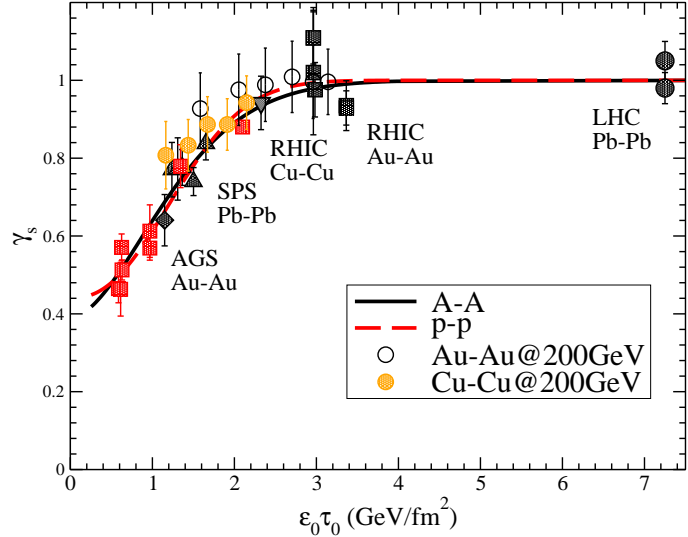}}
\caption{Strangeness suppression as function of the initial energy density
in central $pp$ and $AA$ collisions, as in Fig.\ \ref{supp}; the circles
show the results from non-central collisions, using eq.(15).}
\label{supp-nc}
\end{figure}

As a further test, we can check the variation of $\gamma_s$ with the
centrality of the collision at fixed $A$ and $\sqrt s$. This requires
a definition of the initial energy density for non-central collisions.
In the spirit of our form (\ref{6}), we have used
\be 
\e_0^{N_p}~\tau_0 = {1.5~\! m_T (0.5 N_p) \ \over \pi R_{N_p}^2} 
\left({dN \over dy}
\right)_{y=0}^{AA},
\label{12}
\ee 
with $R_{N_p} = 1.25 (0.5 N_p)^{1/3}$ as an estimate of the energy 
density as function of the number of participants $N_p$. This then
allows us to enter recent data for $\gamma_s$ as function of $N_p$
in $Au-Au$ and $Cu-Cu$ collisions at 200 GeV \cite{central}.
In Fig.\ \ref{supp-nc} it is seen to agree quite well with the
universal curve obtained from the central data. -- In the similar vein,
it would be interesting to see if $pp$ data at fixed $\sqrt s$ also
follow the predicted pattern; for fixed transverse area, $\gamma_s$ 
must increase with multiplicity according to the curve shown in Figs.\ 
\ref{supp} and \ref{supp-nc}, see eq.\ (\ref{8}).

\medskip

We now return to the assumed form ($\ref{6b}$) for the
energy density. In a hydrodynamic description, it is the entropy density
which is related to the number of emitted secondaries,
\be
s_0 = c_s T^3 = {c \over \tau_0 \pi R_A^2} \left({dN \over dy}\right);
\ee
this holds for an ideal gas of massless hadrons at temperature $T$,
with $c_s$ and $c$ determined by the number of degrees of freedom of the
species produced; the transverse nuclear area is given by $\pi R_A^2$.
The corresponding energy density is then given by
\be
\epsilon_0 = (3/4) c_s T^4,  
\ee
so that the relation between energy density and multiplicity is now
given by
\be
\epsilon_0 \tau_0 = (3/4) {\tau_0 \over c_s^{1/3}} 
\left[{c \over \tau_0 \pi R_A^2} 
\left({dN \over dy}\right)\right]^{4/3}. 
\label{x}
\ee
Comparing equ's. (\ref{6b}) and (\ref{x})
we find that as before, when $(dN/dy)(s)$ is given,
$\gamma_s(s)$ and $\epsilon_0(s)$ are
fully specified for a given $s$. Our main result, strangeness
suppression is determined by the initial energy density, thus remains valid. 
More generally speaking, $\gamma_s(s)$ at given $s$ is uniquely correlated 
to the initial energy density at this $s$, as defined in eq. (\ref{6}), 
i.e., to the total (transverse) energy and the transverse area of the 
system.
The specific relation assumed to hold between $dE/dy$ and $dN/dy$ only 
modifies the
detailed form of the scaling variable. In other words, the variation of
$\gamma_s(s)$ with $\epsilon_0 \tau_0$ in Fig.\ \ref{supp}, based on 
eq.\ (\ref{6b}), will coincide with that of $\gamma_s(s)$ as function of
$[(4c_s^{1/3}/3)\epsilon_0\tau_0]^{3/4}$, as given by eq.\ (\ref{x}). 
The basic meaning of the observed universal model-independent 
behavior thus is that $\gamma_s$ is fully determined by the ratio between the 
total transverse energy and the transverse area of the system.

\medskip

It would evidently be interesting to extend these consideration to
hadron production in $e^+e^-$ annihilation. Here the crucial aspect
is a determination of the relevant transverse area; work on this is
in progress.

\medskip

In closing, we recall that our causality considerations relate the size
of the causal correlation region to the life-time of the fireball as
strongly interacting thermal system. Strangeness
suppression provides an experimental measure of the correlation region,
while the initial energy density determines the life-time of the interacting
thermal medium.
The observed scaling of $\gamma_s$ with a function of
$\epsilon_0\tau_0$  is thus an
observable consequence of our basic causality correspondence. 

\medskip

Nevertheless, if one were to just {\sl ad hoc} assume a $\gamma_s(s)-\e_0(s)$ 
correlation, the results of Figs.\ \ref{supp} and \ref{supp-nc} would remain. 
On a purely phenomenological level, one thus also finds that the degree 
of strangeness suppression in hadronic and nuclear collisions is fully 
determined by the initial energy density. A specific example of this
(the $K^- /\pi$ ratio) was already noted some time ago \cite{Wang}.

\vskip1cm

\centerline{\bf \large Acknowledgement}

\bigskip

It is a pleasure to thank J.\ Schukraft for some stimulating questions
and F.\ Karsch for helpful comments.

\end{document}